\documentclass[12pt]{article}
\usepackage{amsmath,amsthm,amssymb,amscd}
\usepackage[english]{babel}
\usepackage{adjustbox}
\usepackage{amsmath,amssymb,amsthm,amsfonts,makeidx,dsfont}
\usepackage{graphicx}
\usepackage{natbib}
\usepackage{color}
\usepackage{lscape} 
\usepackage{lineno} 
\usepackage{mathrsfs}
\usepackage{multirow}

\newtheorem{prop}{Proposition}

\usepackage[left=2cm,right=2cm,top=4cm,bottom=16mm]{geometry}

\begin{document}

\title{Comparing partitions through the Matching Error}

\author{Mathias Bourel\footnote{M. Bourel, mbourel@fing.edu.uy, IMERL, Facultad de Ingenier\'ia, Universidad de la Rep\'ublica, Julio Herrera y Reissig 565, 11300 Montevideo, Uruguay. Corresponding author}, Badih Ghattas\footnote{B. Ghattas, badih.ghattas@univ-amu.fr, Institut de Math\'ematiques de Marseille, Universit\'e Aix-Marseille, France}, Meliza Gonz\'alez \footnote{M. Gonz\'alez, meliza.gonzalez@gmai.com, Universidad de la Rep\'ublica, Montevideo, Uruguay.}}

\maketitle

\begin{abstract} With the aim to propose a non parametric hypothesis test, this paper carries out a study on the Matching Error (ME), a comparison index of two partitions obtained from the same data set, using for example two clustering methods. This index is related to the misclassification error in supervised learning. Some properties of the ME and, especially, its distribution function for the case of two independent partitions are analyzed.  Extensive simulations show the efficiency of the ME and we propose a hypothesis test based on it.  \\

\end{abstract}

\section{Introduction}
  
 Most clustering approaches result in a partition of the data set and often a partition of the space where the data lie. Several indices may be used to compare partitions coming from a same data set, among which the Rand index (\cite{Rand-1971}), the Adjusted Rand Index (\cite{hubert1985comparing}), the Jaccard Index (\cite{hultsch2004untersuchung}), etc. They can be also used to assess the performance of a clustering approach over a supervised dataset. Most of these existing indices lack real mathematical analysis, and almost no information exists about their distribution. 

We consider here the Matching Error (ME) introduced by Meila (\cite{meilua2001experimental}, \cite{meilua2005comparing}) and inspired by the classification error rate used in supervised learning. This index has been used in few works ( for example and recently in \cite{cubt2013}) because its computation for large number of clusters is quite complex. We wish to derive in this paper a hypothesis test to compare two partitions, based on the ME statistic. For that, we focus on the theoretical properties of the ME, in particular to derive its distribution and show  its efficiency in various experimental designs.

This paper is organized as follows. In Section 2, we present a state of the art of some classical indices for comparison of two partitions and their properties. Section 3 is devoted to the study of the ME where we establish some theoretical results, in particular its distribution in case of independence of partitions and balanced clusters. Several properties are proved for the general case. In Section 4, we show many simulations varying the experimental designs (sample size, number of groups, dependence of the partitions) to compare the ME with other indices. This section ends with our proposal of an hypothesis test for the independence of two partitions, designed from the theoretical results on $ME$, and its performance.

\section{Related works}

We denote by $\mathcal{L}=\{x_1,\dots,x_n\}\subseteq \mathds{R}^{p}$ a sample of $n$ independent realizations of a multivariate random variable $X=\left(X_{1},\dots,X_{p}\right)$. Clustering seeks to form disjoint subgroups of observations such that  individuals within the same cluster are similar to each other and relatively different from those of the other clusters. Let  
$\mathscr{C}$ be a partition of $\mathcal{L}$ obtained by a cluster analysis, that is $\mathscr{C}$ is a collection of disjoint subsets $\{C_{1},\dots,C_J\}$ such that their union is $\mathcal{L}$. The set of all possible partitions of  $\mathcal{L}$ is denoted  $\mathcal{P}(\mathcal{L})$. Let  $\mathscr{C}'=\{C'_{1},\dots,C'_{L}\}\in \mathcal{P}(\mathcal{L})$ be a second partition of $\mathcal{L}$. The number of clusters of partitions $\mathscr{C}$ and $\mathscr{C}'$ ($J$ and $L$ respectively) may be different.

Following work of \cite{wagner2007comparing}, there exist three kind of similarity (or dissimilarity) measures between two partitions; we give a brief review of such measures.

\begin{itemize}
\item Measures based on counting pairs 

A natural way to compare partitions is by counting pairs of observations belonging to a same cluster in both partitions. The set of all (unordered) pairs of $\mathcal {L}$ is the disjoint union of the following sets:

\begin{itemize}
\item $A$=\{pairs of observations that are in the same cluster in $\mathscr{C}$ and $\mathscr{C}'$\}
\item $B$=\{pairs of observations that are in different clusters in $\mathscr{C}$ and $\mathscr{C}'$\} 
\item $C$=\{pairs of observations that are in the same cluster in $\mathscr{C}$ but in different clusters in $\mathscr{C}'$\}
\item $D$=\{pairs of observations that are in the same cluster in $\mathscr{C}'$ but in different clusters in $\mathscr{C}$\}
\end{itemize}
  
Sets $A, B, C$ and $D$ are disjoint and if $a=|A|,b=|B|,c=|C|$ and $d=|D|$, where $|\cdot|$ stands for the cardinal) we have $a+b+c+d=\frac{n(n-1)}{2}$.

A very common index based on counting pairs is the \emph{Rand index} (\cite{Rand-1971}) defined by:

	\[R\left(\mathscr{C},\mathscr{C}'\right)=\frac{a+b}{a+b+c+d}=\frac{2(a+b)}{n(n-1)}\]
It counts the proportion of pairs classified in a same way by the two clusterings. It is equal to zero when there exist no pairs of observations classified in the same way by both clustering, and it is equal to one when the two partitions are identical. Because the expected value of the Rand index of two random partitions is not constant, \cite{hubert1985comparing} proposed an adjustment based on the hypothesis that the clusterings are generated randomly subject to a fixed number of groups and fixed cluster size. The \emph{Adjusted Rand} index is a normalized version of the Rand index and is defined as:

\[R_{adj}\left(\mathscr{C},\mathscr{C}'\right)= \frac{R\left(\mathscr{C},\mathscr{C}'\right)-\mathbb{E}(R\left(\mathscr{C},\mathscr{C}'\right))}{\max(R\left(\mathscr{C},\mathscr{C}'\right))-\mathbb{E}(R\left(\mathscr{C},\mathscr{C}'\right))}\]

which is equivalent to: 
\[R_{adj}\left(\mathscr{C},\mathscr{C}'\right)=\frac{a-((a+d)(a+c)/(a+b+c+d))}{\frac{(a+d)+(a+c)}{2}-\frac{(a+d)(a+c)}{a+b+c+d}}\]

Another indices are \emph{Fowlkes-Mallows index} (\cite{fowlkes1983method}), \emph{Mirkin metrix}(\cite{mirkin1998mathematical}), \emph{Partition Difference} (\cite{li2004combining}) and
\emph{Jaccard index} (\cite{hultsch2004untersuchung}) . The latter measures the similarity between two partitions. It is very similar to the Rand index, but it dismisses the pairs of elements that are in different clusters in the compared partitions. It is defined as

	\[J(\mathscr{C},\mathscr{C}')=\frac{a}{a+c+d}\]
 
However, many of these measures have undesirable properties such as sensitivity to the number of clusters, the number of observations and the relative size of clusters. 

\item Measure based on set overlaps

Measurements based on  set overlaps are in general computed from the confusion matrix between the two partitions $\mathscr{C} $ and $ \mathscr{C}'$. The matrix  $ N = (n_ {ij }) \in J \times L $ is such that $ n_{ij} = \left| C_{i} \cap C_{j}^{'} \right| $, $ 1 \leq i \leq J $, $ 1 \leq j \leq L $. We will suppose that $J \leq L$.

\cite{meilua2001experimental} introduced an index called \emph{the classification error} inspired from the misclassification error used in supervised learning. Consider that one of the two compared clusterings ($\mathscr{C}$ for instance) corresponds to the true labels of each observation and the other clustering ($\mathscr{C}'$) to the predicted ones. The supervised classification error may be computed for all the possible permutations of the predicted labels (in $\mathscr{C}'$), and the maximum error over all the permutations may be taken. Thus the classification error for comparing both partitions may be written as

\begin{equation} 
CE(\mathscr{C},\mathscr{C}')=1-\frac{1}{n} \underset{\sigma}{\max} \sum \limits_{i=1}^{J} n_{i \sigma(i)}
\label{mce1}
\end{equation}

where $\sigma$ is an injective mapping of  $\{1,\dots,J\}$ into $\{1,\dots,L\}$ (\cite{meilua2005comparing}).
The $ME$ index may be complex to compute when the number of clusters is large. A polynomial time algorithm has been proposed by \cite{cubt2013} to compute it efficiently. We will study the distributional properties of this index in the next section.

\item Measures based on mutual information

The entropy of a partition $\mathscr{C}$ is defined by $H(\mathscr{C})=-\sum \limits^{J}_{i=1}p(i)\log_{2}p(i)$ where $p(i)=\left|C_{i}\right|/n$ is the estimate of the probability that an element is in cluster $C_{i} \in \mathscr{C}$.
The \emph{mutual information} can be used to measure the independence of two partitions $\mathscr{C}$ and $\mathscr{C}'$. It is given by: 	
$I(\mathscr{C},\mathscr{C}')=\sum\limits^{J}_{i=1} \sum \limits^{L}_{j=1}p(i,j)\log_{2}\frac{p(i,j)}{p(i)p(j)}$, where $p(i,j)$ is the estimate of the probability that an element belongs to cluster $C_i$ of $\mathscr{C}$ and $C'_j$ of $\mathscr{C}'$. Mutual information is a metric over the space of all clusterings, but its value is not bounded which makes it difficult to interpret. As $I(\mathscr{C},\mathscr{C}')\leq \min\left(H(\mathscr{C}),H(\mathscr{C'})\right)$, other bounded indices have been proposed such as \emph{Normalized Mutual Information} (\cite{strehl2002}, \cite{fred2003}) where $I(\mathscr{C},\mathscr{C}')$ is divided either by the arithmetic or the geometric mean of the clustering entropies. Meila (\cite{meila2003varinf}) has also proposed an index based on Mutual information called \emph{Variation of Information}.
\end{itemize}

In \cite{de2012comparison}  and \cite{rezaei2016set} several indices are compared on artificially simulated partitions with various configurations; partitions are either balanced or unbalanced, dependent or independent, varying number of clusters. They show that the indices based on set overlaps have better performance than those based on counting pairs and mutual information. Besides, most indices are not relevant when the clusters in the partitions are imbalanced.\\ 

\indent \cite{milligan1986study} study the behavior of the Rand, Adjusted Rand, Jaccard and Fowlkes Mallows indices. They compare the partitions produced by hierarchical algorithms with the true partitions, varying the number of groups with a sample of 50 observations and conclude that the adjusted Rand index seems to be more appropriate for clustering validation in this context. Similar simulations and results are given in \cite{brun2007model}  and \cite{wu2009external} using $k$-means. \\

The works cited above give some experimental conclusions with no theoretical framework. In \cite{saporta2002comparing}, \cite{youness2004some} and \cite{youness2010comparing}, the authors propose  methods to study the empirical distribution of partition comparison indices, in particular the Rand and the Jaccard indices, among others. By a latent class mode to generate the data, they estimate their empirical distribution under the hypothesis that the two partitions come from the same underlying mixture model. Distributions of these indices depend on the number of clusters, their proportions and their separation so it is impossible to derive a general result.
 
\vspace{2mm}
Indices for comparing partitions should have some desirable properties, like being bounded, interpretable, independent of the number of clusters and sample size, and complying with properties of a metric. 
In \cite{meilua2005comparing}, the author makes an axiomatic characterization of Variation of information, Mirkin, Rand and Van Dongen indices and some properties of the ME are discussed, in particular that it is a metric in some subspaces of the clusterings sets. In \cite{meilua2007comparing} numerous theoretical properties on the variation of information index are proved. Also, it is shown that the normalized Mirkin metric and the Adjusted Rand index satisfy the properties of a metric.

With respect to the theoretical distribution of the indices, \cite{idrissi2000contribution} established the following statement: for two independent partitions, with $J$ balanced clusters, the asymptotic distribution of the Rand Index is normal with expectation $\mathds{E} \left (R \left(\mathscr{C}, \mathscr{C}'\right) \right) = 1- \frac{2}{J} + \frac{2}{J^{2}}$ and variance $ \mathds{V} \left(R \left (\mathscr{C}, \mathscr{C}'\right) \right) = \frac{1}{n^{2}} \left (1- \frac{1}{n} \right) \left (1- \frac{2}{J} + \frac {2}{J^{2}} \right) \left (\frac{2}{J} - \frac{2}{J^{2}} \right)$. But this result is not true for small $J$, in particular when $J=2$, and is only approximately valid only for large samples. This is, to our best knowledge, the only theoretical result about the distribution of the Rand index and under some conditions. 
Our work intends to analyze the distribution of the $ME$ index.

\section{The ME properties}

We now study the distributional properties of the classification error introduced in (\ref{mce1}) that we formulate in an equivalent way. More precisely, let us consider a data set $\mathcal{L} = \left\{x_{1},\dots,x_{n}\right\}\subseteq \mathds{R}^{p}$ and $\mathscr{C}$ and $\mathscr{C}'\in \mathcal{P}(\mathcal{L})$ two partitions of $\mathcal{L}$. The labels of each observation in the first partition are denoted $\left\{y_{1},\dots,y_{n}\right\}$ and those of the second partition $\left\{\hat{y}_{1},\dots,\hat{y}_{n}\right\}$, so $y_{i}\in\left\{1,\dots,J\right\}$ and $\hat{y_{i}}\in\left\{1,\dots,L\right\} \forall i=1,\dots,n$.\par

If $S_{J}$ is the set of permutations of $\left\{1,\dots,J\right\}$ and $\mathcal{A}$ is the set of arrangements of $J$ elements taken from $L$, we define the \emph{Matching Error} (ME) as: 

 \begin{equation} \tau= ME(\mathscr{C},\mathscr{C}') =\left\{
	       \begin{array}{ll}
		\underset{\sigma \in S_{J}}{\min} \frac{1}{n}\sum \limits^{n}_{i=1}\mathbf{1}_{\{{y_{i}}\neq\sigma(\hat{y}_i)\}}   & \mathrm{if}\, J\leq L \\
		 \underset{\sigma \in \mathcal{A}}{\min} \frac{1}{n}\sum \limits^{n}_{i=1}\mathbf{1}_{\{\sigma(y_{i})\neq\hat{y}_i\}} & \mathrm{otherwise\ } \\
		    \end{array}
	    \right.\end{equation}

For simplicity, we assume that the two compared partitions have the same number of clusters, that is $J = L$. Observe that $ME(\mathscr{C},\mathscr{C}')$ is another formulation of $CE(\mathscr{C},\mathscr{C}')$ used in the works of Meil{\u{a}} and we are going to analyze its distribution.

\subsection{Distribution function of $\tau_\sigma$}

Let denote $\tau_{\sigma}=\frac{1}{n}\sum \limits^{n}_{i=1}\mathbf{1}_{\{{y_{i}}\neq\sigma(\hat{y}_i)\}}$. To derive the distribution of the $ME$ we need to know the distribution of the different classification errors $\tau_{\sigma}$ where $\sigma \in S_J$. For $y_{i}$ and $\hat{y_{i}}$ two independent realizations of a discrete random variable $Y$ taking values in $\{1,\dots,J\}$, let:

\begin{itemize}
\item  $p_j=\mathds{P}[y_{i}=j]$, $\hat{p_j}=\mathds{P}[\hat{y}_i=j]$, and
\item$\theta=\mathds{P}[y_{i}\neq\sigma(\hat{y}_i)]=1-\mathds{P}[y_{i}=\sigma(\hat{y}_i)]\\ 
=1-\sum \limits^{J}_{j=1}\mathds{P}[y_{i}=j,\sigma(\hat{y}_{i})=j]=1-\sum \limits_{j=1}^J p_{j}\hat{p}_{j}$ 
\end{itemize}

Assuming that $Y$ is uniform on $\left\{1,\dots,J\right\}$, then $p_{j}=\hat{p}_{j}=1/J$ and $\theta=1-\frac{1}{J}$. Therefore the random variable $n\tau_{\sigma}$ is binomial with parameters $n$ and $\theta$, and:
$$
	\mathds{E}(\tau_{\sigma})=1-\frac{1}{J};\,\,\,\,\,\textrm{and}\,\,\,\,\,\,\,\mathds{V}(\tau_{\sigma})=\frac{1}{n}\frac{1}{J}\left(1-\frac{1}{J}\right)$$

\noindent For large values of $n$: $n\tau_{\sigma}\sim \mathcal{N}\left(n\theta,n\theta(1-\theta)\right)$.

\begin{prop}
\begin{enumerate}

\item As $\sum\limits^{J!}_{j=1}n\tau_{\sigma_{j}}=n(J-1)!(J-1)$, the random  variables 
$n\tau_{\sigma_{1}},n\tau_{\sigma_{2}},\dots, n\tau_{\sigma_{J!}}$ are not independent.
\item The ME is bounded: $$
0\leq\tau\leq\frac{J-1}{J}$$
\end{enumerate}

\end{prop}

Let $\sigma_l$ and $\sigma_k$ be two permutations of $S_J$. We say that $\sigma_l$ and $\sigma_k$ share a point $j \in \{1,\dots,J\}$ if $\sigma_l(j)=\sigma_k(j)$. Note that two permutations of $S_J$ can share at most $J-2$ points.

\begin{prop}
If $\sigma_j$ and $\sigma_l$ share $s$ points, then
$$\mathds{COR}(n \tau_{\sigma_l}, n \tau_{\sigma_k})=\frac{s-1}{J-1}\,\, $$

where $s=0,\dots,J-2$
\end{prop}

\begin{figure}[!ht]
    \centering
   \includegraphics[scale=.3]{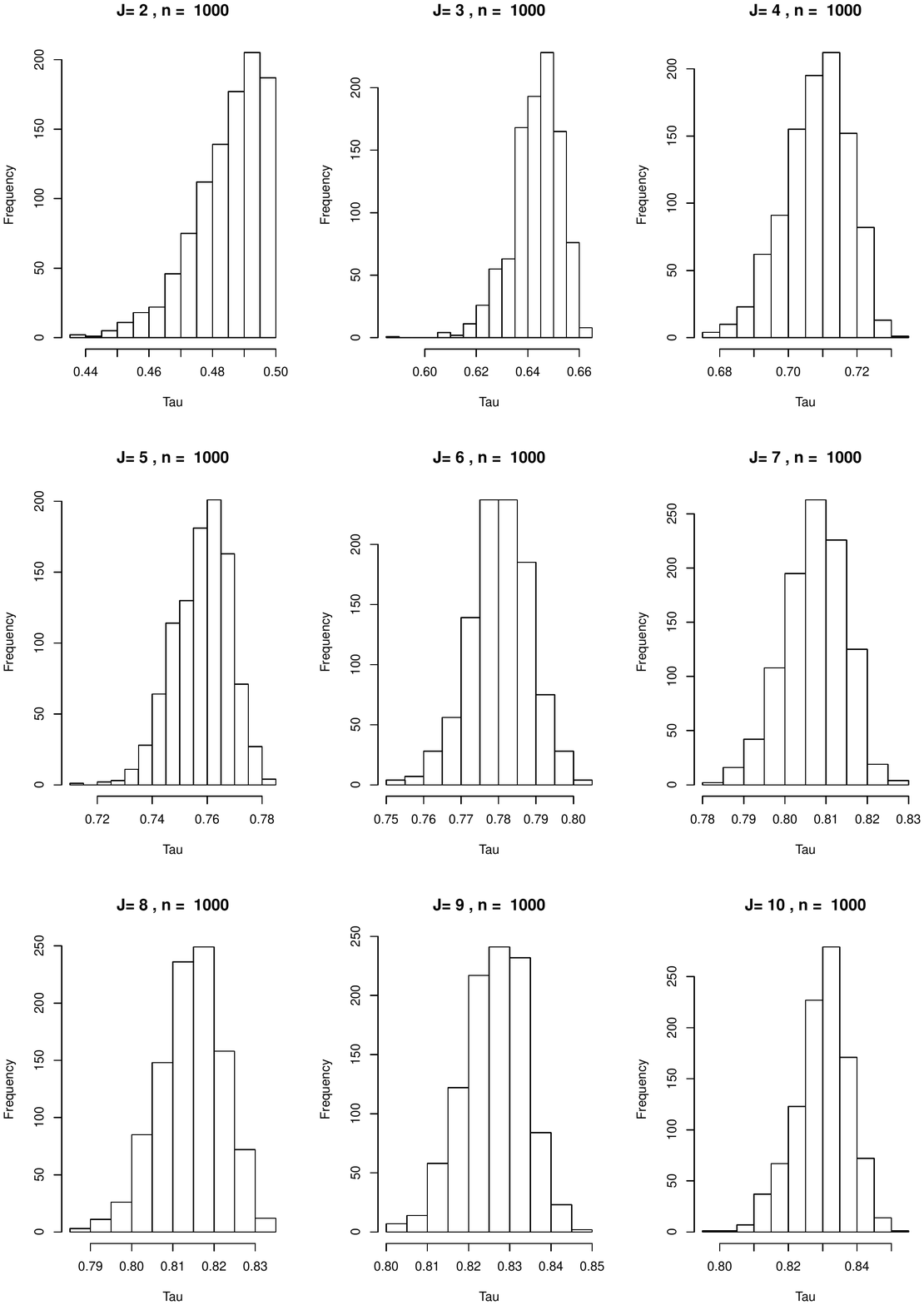}
    \caption{Distribution of $\tau$ for $n=1000$ and different values of $J$ for two independent and balanced partitions.}
    \label{fig:dist_tau}
\end{figure}

In Figure \ref{fig:dist_tau} we show the distribution of $\tau$ for different values of $J$ and for $n=1000$.

\subsection{The case $J=2$}

\begin{prop}
The distribution function $F_{n\tau}(z)$ of $n\tau=\min\left\{n\tau_{\sigma_{1}},n\tau_{\sigma_{2}}\right\}$ is:

$$	
	 \mathds{P}(n\tau \leq z)= \left\{
	       \begin{array}{ll}
		 \sum \limits^{z}_{i=0} 2\binom{n}{i} \left(\frac{1}{2}\right)^{n}= 2 \mathcal{B}_{n,1/2}(z)	& \mathrm{if}\, z+1\leq \frac{n}{2}\\
		  1      & \mathrm{if\ } z>\frac{n}{2} \\
		    \end{array}
	     \right.
$$

where $B_{n,1/2}$ is the binomial distribution function with parameters ($n$, $\frac{1}{2}$).  
\end{prop}

\begin{prop}
\[ 
	\mathds{E}(n\tau)= \left\{
	       \begin{array}{ll}
		 \frac{n}{2}-\left(\frac{1}{2}\right)^{n}\binom{n}{n/2}\frac{n}{2}	& \mathrm{if\ } n \mathrm{\ } \mathrm{is\ } \mathrm{even\ }   \\
		  \frac{n}{2}-\left(\frac{1}{2}\right)^{n}\binom{n-1}{\frac{n-1}{2}}n & \mathrm{if\ } n \mathrm{\ } \mathrm{is\ } \mathrm{odd\ }  \\
		    \end{array}
	     \right.\]
\end{prop}

Hence the distribution of $ME$ is explicit for $J=2$.

\section{Simulations and hypothesis test}	
We first show how the ME behaves in different experimental situations and derive a hypothesis test for independence of two partitions.

\subsection{Empirical performance of the ME}
In this section we study the empirical distribution of $ME$ between two partitions $\mathscr{C}$ and $\mathscr{C}'$. The results are considered and compared in various  experimental conditions: different number of groups ($J\in\left\{2,3,4,5,6,7,8,9,10\right\}$), different number of observations ($n\in \left\{50,100,200,300,400,500,1000\right\}$), independent partitions, different degrees of dependence between partitions, balanced and imbalanced clusters. \\
For each configuration of these parameters, we directly generate the partition vectors, that is, each observation's label for both partitions. These could be the result of two clustering analysis. For each value of $J$ and $n$, we generate $N =1000$ independent partition pairs and the index $\tau = ME (\mathscr{C}, \mathscr{C}')$ is computed. This produces $1000$ values of the index for each configuration. 

To simulate dependent partitions, we start with two equal vectors $Y$ and $\hat{Y}$ and modify at random a proportion $\gamma \in \{0{.}1,0{.}4,0{.6},0{.}9\}$ of labels of $\hat{Y}$. \\
We check first the correlations between the $ME$, and both the Rand and Jaccard indices. Tables \ref{tab:re} and \ref{tab:re2} give the obtained results for the four scenarios, for $\gamma=0.4$.

\begin{table*}[!t]
\centering
\caption{Correlation between $ME(\mathscr{C},\mathscr{C}')$ and $R(\mathscr{C},\mathscr{C}')$ over $N=1000$ repetitions of the indices, where $\mathscr{C}$ and $\mathscr{C}'$ are partitions with $J=2,\dots,10$ groups and $n=1000$ observations in the four scenarios explained above.}
\label{tab:re}
\resizebox{15cm}{!} {
\begin{tabular}{rrcccccccccc}
  \hline
 & & J=2 & J=3 & J=4 & J=5 & J=6 & J=7 & J=8 & J=9 & J=10 & mean\\ 
 \hline
\multirow{ 2}{*}{Dependence}&	Unbalanced groups &-1&	-0,93&	-0,89&	-0,83&	-0,81&	-0,79&	-0,73&	-0,74&	-0,70 &	-0,83 \\
&	Balanced groups&-1	&-1	&-1	&-1	&-0,99	&-0,99	&-0,99	&-0,99	&-0,98	&-0,99\\
  \hline
	\multirow{ 2}{*}{Independence}&	Unbalanced groups&-0,94	&-0,28&	-0,43&	-0,21&	-0,21	&-0,12&	-0,14&	-0,13	&-0,05	&-0,28 \\
&Balanced groups 	&-0,93	&-0,86	&-0,83	&-0,81	&-0,8&	-0,79&	-0,78&	-0,78&	-0,77&	-0,82\\
 	\hline\end{tabular}
	}

\end{table*}

\begin{table*}[!t]
\centering
\caption{Correlation between $ME(\mathscr{C},\mathscr{C}')$ and Jaccard$(\mathscr{C},\mathscr{C}')$ over $N=1000$ repetitions of the indices, where $\mathscr{C}$ and $\mathscr{C}'$ are partitions with $J=2,\dots,10$ groups and $n=1000$ observations in the four scenarios explained above.}
\label{tab:re2}
\resizebox{15cm}{!} {
\begin{tabular}{rrcccccccccc}
  \hline
 & & J=2 & J=3 & J=4 & J=5 & J=6 & J=7 & J=8 & J=9 & J=10 & mean\\ 
 \hline
\multirow{ 2}{*}{Dependence}&	Unbalanced groups  &		-0,99	&-0,95	&-0,91	&-0,87&	-0,85&	-0,82&	-0,78&	-0,77	&-0,75	&-0,86 \\
&	Balanced groups	&-1	&-1	&-1	&-1	&-1	&-1	&-1	&-1	&-1&	-0,99\\
  \hline
	\multirow{ 2}{*}{Independence}&	Unbalanced groups  &-0,58	&-0,45	&-0,6	&-0,52	&-0,56	&-0,59	&-0,57	&-0,63	&-0,54	&-0,56 \\
&	Balanced groups	&-0,93&	-0,86&	-0,83&	-0,81&	-0,8&	-0,79&	-0,78&	-0,78	&-0,77&	-0,82\\
 	\hline\end{tabular}
	}

\end{table*}

In the case of balanced clusters, the correlation is high (it is negative as $ME$ is a dissimilarity measure between partitions, whereas Rand and Jaccard indices are similarity measures). For dependent partitions the correlation is almost perfect (close to minus one), while with independent partitions it is still high but smaller (between $ -0.77 $ and $ -0.93 $). Correlation between the indices decreases when the number of clusters $J$ increases.\\
For unbalanced partitions results are quite different. For dependent partitions, the correlations -although  decreasing with respect to the cluster size- have a coefficient between $ -0.7 $ and $ -0.99 $. However, when the partitions are independent, the correlation is  smaller and, except for the case $ J = 2 $, it is much weaker with the Rand index than with the Jaccard index.

\subsection{Hypothesis Test}

Our main purpose when analysing the distribution of the $ME$ index is to design a hypothesis test to decide whether two partitions are statistically independent. The properties proved above were derived under some assumptions and may be used to compare partitions at least for $ J = 2 $ groups equally distributed. We present this test and analyse its performance on simulated data. 

Given two partitions $\mathscr{C}$ and $\mathscr{C}'$, the test proposal is: 

$(H_0)$: Partitions $\mathscr{C}$ and $\mathscr{C}'$ are independent.

$(H_1)$: Partitions $\mathscr{C}$ and $\mathscr{C}'$ are not independent.

The test statistic is $\tau$ and, as we know its distribution for $J=2$ under $H_0$ it is straightforward to derive decision. 

\begin{figure}[!ht]
    \centering
    \includegraphics[scale=.2]{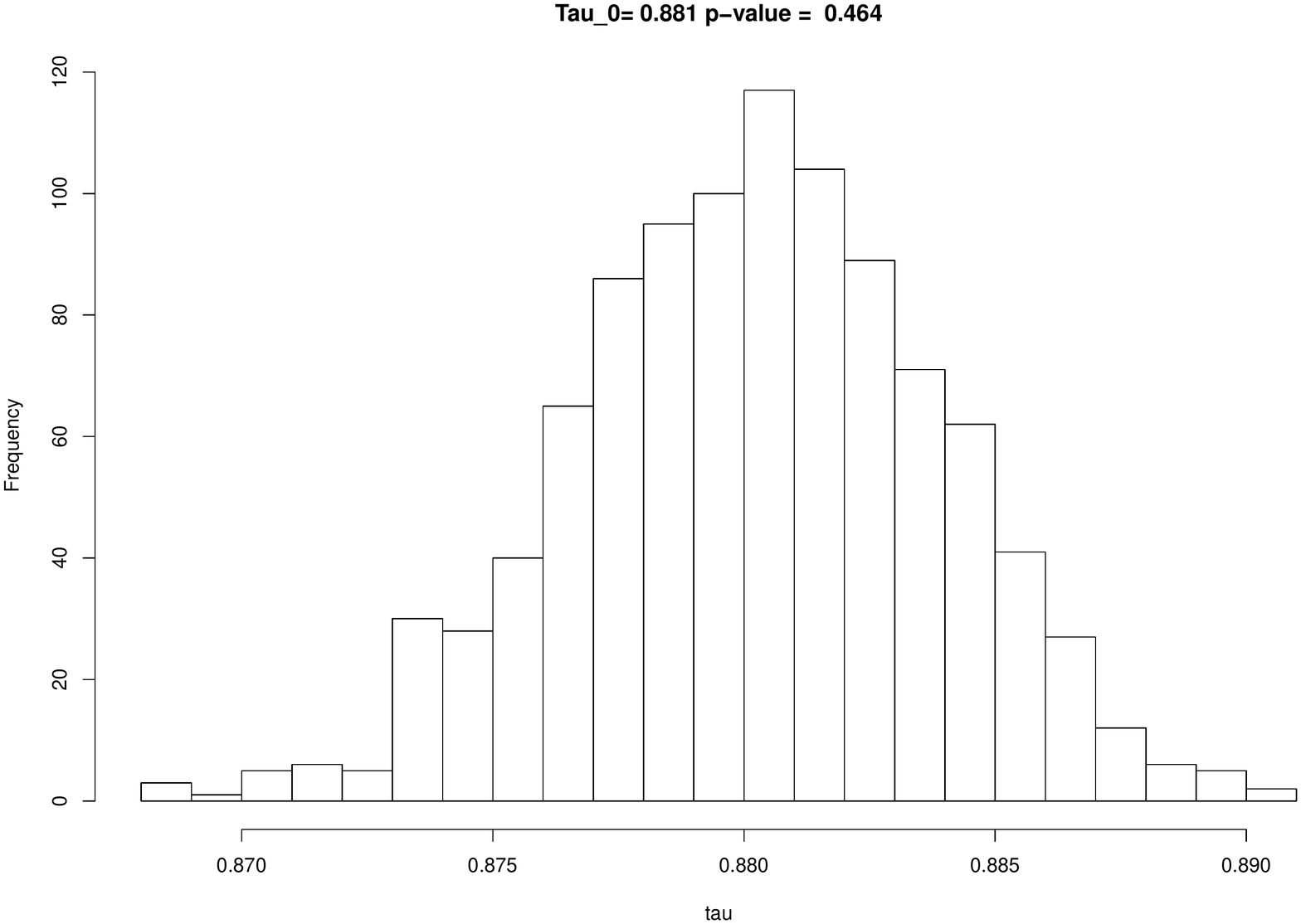}
    \caption{Histogram of the $B=1000$ values of $\tau$ obtained by changing $20\%$ of the labels randomly with two independent partitions.}
    \label{fig:ind}
%
    \includegraphics[scale=.2]{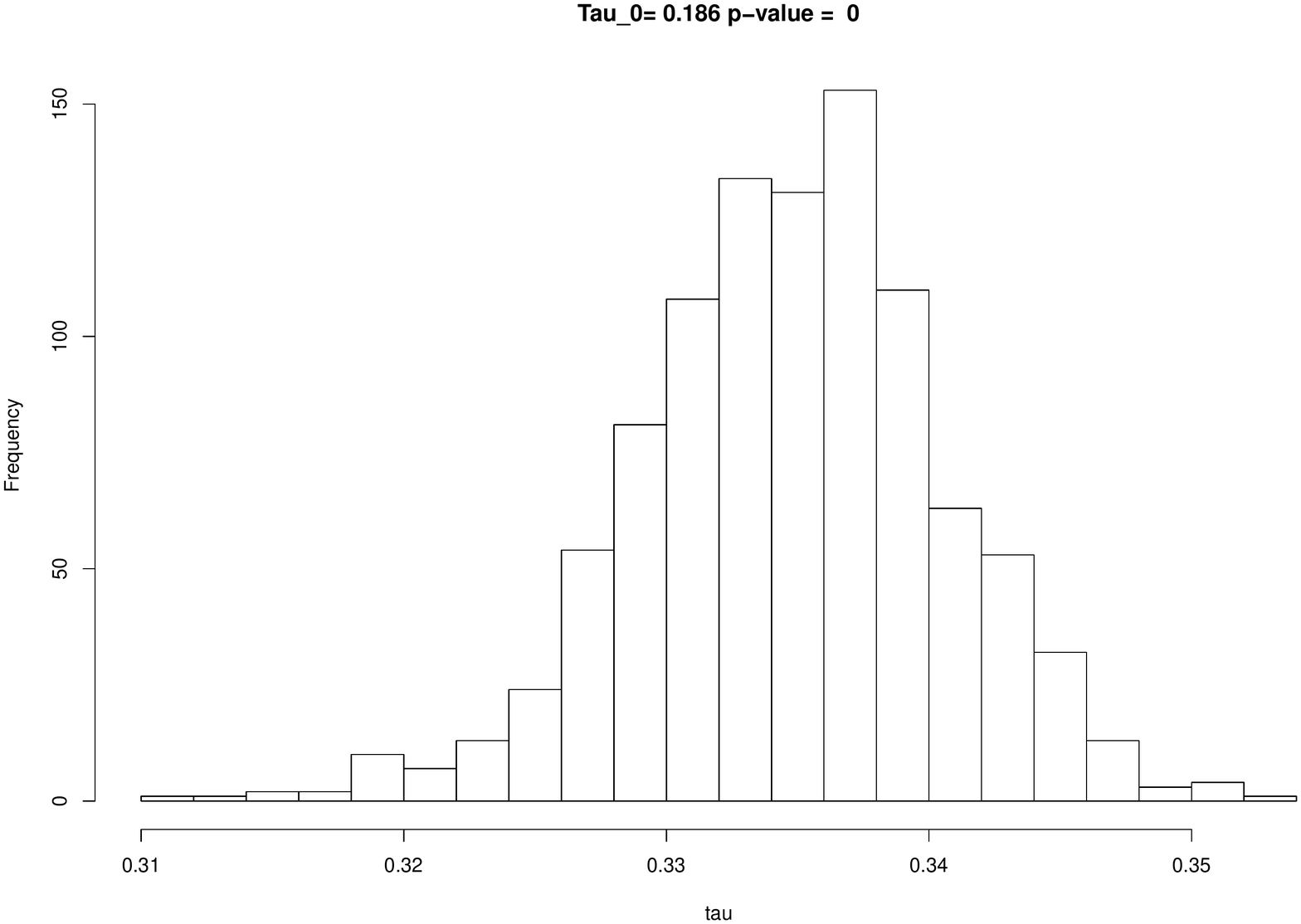}
    \caption{Histogram of the $B=1000$ values of $\tau$ with two dependent partitions.}
    \label{fig:dep}
\end{figure}

For $J>2$, we propose the following reasoning. For two partitions $\mathscr{C}$ and $\mathscr{C}'$ we compute $\tau_0=ME(\mathscr{C},\mathscr{C}')$ and then take $B$ ``perturbations'' of $\mathscr{C}'$ denoted $\mathscr{C}'_1, \mathscr{C}'_2, \dots, \mathscr{C}'_B,$ changing at random a proportion $\pi$ of its labels. To estimate the $p$-value $\mathds{P}(\tau \leq \tau_0)$ of the independence hypothesis (null hypothesis) we take the proportion of values of $\tau_b=ME(\mathscr{C},\mathscr{C}'_b)$ which are less than $\tau_0$.  We give two examples of the use of this test, for $J=15$. First we consider two independent balanced partitions of $n=1000$ observations with $J=15$ groups and fix $B=1000$. In this case $\tau_0=0{.}881$ and the estimated $p-$value is $0{.}464$, which is coherent with not rejecting the hypothesis of independence of the two partitions. In Figure \ref{fig:ind} we show the histogram of the $B$ different values of $\tau$. In the second case and for the same values of $n$ and $J$, we consider two dependent partitions where the second is constructed from the first by changing $20\%$ of the labels randomly. In this case $\tau_0=0{.}186$ and the estimated $p$-value equals 0 so we reject the hypothesis of dependence. In Figure \ref{fig:dep} we show the histogram of the $B$ different values of $\tau$.

To evaluate the performance of the test, we calculate the error averaging over $N=1000$ simulations. We take $B=1000$ and proportion $\pi$ equals $0{.}2$. At level $\alpha=0{.}1$, we estimate the number of times we make type I and type II error using the empirical values of the $p$-value. As it was expected, taking proportion of values of $p$-value less than $0.1$, the estimation of doing a type I error is small, less than $8\%$, when we compare two independent partitions. We give in Table \ref{tab:tipo1} the individual values for some $n$ and varying $J\in \{3,\dots,10\}$ .

\begin{table}[ht]
\centering
	\caption{At level $0.1$, proportion of times a type I error is made obtained by averaging $N =1000$ comparisons of independent partitions with balanced classes.}\label{tab:tipo1}
\begin{tabular}{rrrrrrrrrr}
  \hline
  & J=3 & J=4 & J=5 & J=6 & J=7 & J=8 & J=9 & J=10 \\ 
  \hline
n=50  & 1.1 & 1.6 & 3.3 & 3.1 & 4.3 & 6.1 & 7.9 & 7.1 \\ 
  n=100  & 0.3 & 1.2 & 1.6 & 3.1 & 3.6 & 3.1 & 4.0 & 4.5 \\ 
  n=200  & 0.2 & 0.8 & 1.2 & 1.5 & 1.7 & 2.5 & 2.8 & 3.0 \\ 
  n=300  & 0.4 & 0.7 & 1.1 & 1.3 & 1.6 & 1.1 & 2.4 & 2.3 \\ 
  n=400  & 0.2 & 0.4 & 0.6 & 1.3 & 1.8 & 1.3 & 1.4 & 1.7 \\ 
  n=500  & 0.0 & 0.2 & 1.1 & 0.9 & 1.2 & 1.7 & 1.4 & 2.8 \\ 
  n=1000 & 0.0 & 0.1 & 0.8 & 0.9 & 0.8 & 1.4 & 1.4 & 2.2 \\ 
   \hline
\end{tabular}

\end{table}

On the other hand, with dependent partitions, to estimate the proportion of type II errors over the $N$ simulations, we take the proportion of $p$-values larger than $0.1$. As described in the previous section, partitions with balanced clusters and different degrees of dependence are simulated where dependency strength $\gamma$ varies in $\left \{0.1, 0.4, 0.6, 0.9 \right\}$.  This estimation equals zero when $\gamma=0.1$ and $\gamma=0.4$ accordingly with a high dependence of the partitions. For $\gamma=0.6$ it is null except for $n=50$: for $J=3$ it equals $17{.}8\%$, and it is less than $3\%$ only for $4 \leq J \leq 10$. When $\gamma=0{.}9$, which is a scenario very close to independence, it has high values for $n=50,100$ an $200$  but decreases to $0$ when $n$ and $J$ grow as we can see in Table \ref{tab:e2_9}.

\begin{table}[ht]
\centering
\caption{At level $0.1$, proportion of times a type II error is made obtained averaging $N =1000$ comparisons of dependent partitions with $\gamma=0{.}9$, with balanced classes.}
\label{tab:e2_9}
\begin{tabular}{rrrrrrrrrr}
  \hline
 & J=3 & J=4 & J=5 & J=6 & J=7 & J=8 & J=9 & J=10 \\ 
  \hline
n=50 &  98.3 & 97.4 & 96.0 & 95.3 & 94.4 & 93.3 & 92.2 & 91.9 \\ 
  n=100  & 97.7 & 96.4 & 95.1 & 93.6 & 91.1 & 91.3 & 90.4 & 91.8 \\ 
  n=200  & 94.3 & 89.0 & 87.7 & 83.0 & 83.1 & 82.5 & 79.2 & 82.8 \\ 
  n=300  & 89.5 & 78.9 & 72.7 & 65.9 & 67.5 & 63.9 & 63.2 & 62.7 \\ 
  n=400  & 81.8 & 63.0 & 53.4 & 46.1 & 43.1 & 40.8 & 40.5 & 38.1 \\ 
  n=500  & 73.1 & 46.5 & 35.0 & 27.9 & 21.5 & 23.4 & 20.3 & 17.3 \\ 
  n=1000 & 22.2 & 3.7 & 1.6 & 0.2 & 0.2 & 0.0 & 0.0 & 0.0 \\ 
   \hline
\end{tabular}

\end{table}

The experimental evaluation shows a good performance of the partition independence test with balanced clusters. The proportion of times for which a type I error occurs is in the expected environment for a level test with $\alpha = 0.1$. In the evaluation of type II error, the test proved to be powerful when the dependence between the partitions is relatively high and that it loses precision when the compared partitions have conditions that approximate the null hypothesis of independence.

\section{Conclusions}
We have suggested an hypothesis test for comparing two partitions, useful for comparing the results of two clustering approaches over a same dataset. Our test is based on the mismatch error inspired from the misclassification error in supervised learning. We have analyzed the properties and the distribution of this index in several conditions and compared it to other common indices. A closed form of the statistic distribution under the null hypothesis was given for two clusters under mild conditions. For more than two clusters, the simulations show that the test is quite robust and reliable in various experimental conditions, but the statistic distribution under the null hypothesis is still unavailable.

\section*{Appendix}

\paragraph{Proof of Proposition 1}

\begin{enumerate}
    \item
\begin{flushleft}
$n\sum\limits^{J!}_{j=1}\tau_{\sigma_{j}}=n\frac{1}{n}\sum\limits^{n}_{i=1}\left(\mathbf{1}_{\{y_{i}\neq \sigma_{1}(\hat{y}_{i})\}}+\dots+\mathbf{1}_{\{y_{i}\neq \sigma_{J!}(\hat{y}_{i})\}}\right)=\sum\limits^{n}_{i=1}\left|\left\{\sigma \in S_{J}: \sigma(\hat{y_{i}})\neq y_{i}\right\}\right|$\\
\end{flushleft}
For fixed $y_{i}$ and $\hat{y_{i}}$  we have: $\left|\left\{\sigma \in S_{J}: \sigma(a)\neq b\right\}\right|=$\\ $\left|S_{J}\right|- \left|\left\{\sigma \in S_{J}: \sigma(a)=b\right\}\right|=J!-(J-1)!=(J-1)!(J-1)$\\
it follows that $\,\,\,n\sum\limits^{J!}_{j=1}\tau_{\sigma_{j}}=n(J-1)!(J-1)$.

\item 
 It is straightforward, because if $\tau$ was greater than $\frac{J-1}{J}$, then $\sum\limits^{J!}_{j=1}\tau_{\sigma_{j}}>J!\frac{J-1}{J}=(J-1)!(J-1)$, which contradicts part 1.

\end{enumerate}

\paragraph{Proof of Proposition 2}
 \begin{align*}
& \mathds{COV}(n\tau_{\sigma_{k}},n\tau_{\sigma_{l}})=\mathds{E}(n\tau_{\sigma_{k}},n\tau_{\sigma_{l}})-\mathds{E}(n\tau_{\sigma_{k}})\mathds{E}(n\tau_{\sigma_{l}})\\
=&\mathds{E}\left(\sum\limits^{n}_{i=1}\mathbf{1}_{\{y_{i}\neq \sigma_{k}(\hat{y}_{i})\}}\sum\limits^{n}_{j=1}\mathbf{1}_{\{y_{j}\neq \sigma_{l}(\hat{y}_{j})\}}\right)-\mathds{E}\left(\sum\limits^{n}_{i=1}\mathbf{1}_{\{y_{i}\neq \sigma_{k}(\hat{y}_{i})\}}\right)\mathds{E}\left(\sum\limits^{n}_{j=1}\mathbf{1}_{\{y_{j}\neq \sigma_{l}(\hat{y}_{j})\}}\right)\\
=& \sum\limits^{n}_{i=1}\sum\limits^{n}_{j=1} \mathds{P}\bigl(y_{i}\neq \sigma_k(\hat{y}_{i}), y_{j} \neq \sigma_l(\hat{y}_{j})\bigr)-n^2\bigl(1-1/J\bigr)^2
\end{align*}
Using De Morgan equalities and basic probability properties, it is easy to show that, by independence $\mathds{P}\left(y_{i}= \sigma_k(\hat{y}_{i}),y_{j}= \sigma_l(\hat{y}_{j})\right)$ 
\begin{align*}
    = \left\{\begin{array}{l} 1/J^2\,\, \textrm{if}\,i\neq j \\  
    \sum \limits_{j=1}^{J}\mathds{P}\left(y_{i}=j\right)\underbrace{\mathds{P}\left( \sigma_k(\hat{y}_{i})=j,\sigma_l(\hat{y}_{i})=j\right)}_{\left(\ast\right)}\,\, \textrm{if}\,i= j\end{array}\right.
\end{align*}

If $\sigma_k$ and $\sigma_l$ share $s$ points, $(\ast)=\frac{s}{J^2}$ and therefore 
$$\mathds{COV}(n\tau_{\sigma_k}, n \tau_{\sigma_l})=\frac{n(s-1)}{J^2}$$ and 
$$\mathds{COR}(n\tau_{\sigma_k}, n \tau_{\sigma_l})=\frac{\frac{n(s-1)}{J^2}}{n\frac{1}{J}\left(1-\frac{1}{J}\right)}=\frac{s-1}{J-1}$$

\paragraph{Proof of Proposition 3}
As $n\tau_{\sigma_{1}}+n\tau_{\sigma_{2}}=n$, if $n\tau_{\sigma_{1}}
=x$ and $n\tau_{\sigma_ {2}}=y$, it is easy to establish that: 

$$
\mathds{P}(n\tau_{\sigma_{1}}=x,n\tau_{\sigma_{2}}=y)= \left\{
	       \begin{array}{ll}
		 	0 & \mathrm{if\ }  x+y\neq n \\ \\
		 \binom{n}{x}\left(\frac{1}{2}\right)^{n}   & \mathrm{if\ } x+y=n \\
		    \end{array}
	     \right.
$$

\noindent and the joint probability table for $n\tau_{\sigma_{1}}$ and $n\tau_{\sigma_{2}}$ has the following shape:

	\[ \resizebox{8.5cm}{!}{$\displaystyle \begin{array}{c|ccccccc}
n\tau_{\sigma_{1}}/n\tau_{\sigma_{2}} &0 &1 &2&\cdots &n-2&n-1&n\\
 \hline
0&0 &0 &0 & 0&0&0&\binom{n}{0} \left(\frac{1}{2}\right)^{n}\\
1& 0&0 &0&0&0&\binom{n}{1} \left(\frac{1}{2}\right)^{n}& 0\\
2&0 &0&0&0&\binom{n}{2} \left(\frac{1}{2}\right)^{n} &0 & 0\\
\vdots&0 &0&0&\cdots&0&0 & 0\\
n-2&0 &0&\binom{n}{n-2}\left(\frac{1}{2}\right)^{n} &0&0 & 0& 0\\
n-1&0 &\binom{n}{n-1}\left(\frac{1}{2}\right)^{n} &0 &0&0& 0& 0\\
n&\binom{n}{n} \left(\frac{1}{2}\right)^{n}&0 &0 &0& 0&0& 0\\
 \end{array} $}.\]

Using this table, we have;
\begin{align*}
\mathds{P}(n\tau = z) & \\
=& \mathds{P}(n\tau_{\sigma_{1}}= z,n\tau_{\sigma_{2}}= n-z)+\mathds{P}(n\tau_{\sigma_{2}}= z,n\tau_{\sigma_{1}}= n-z)\\
= &  2\mathds{P}(n\tau_{\sigma_{1}}= z,\, n\tau_{\sigma_{2}}= n-z)\\ 
= & 2\binom{n}{z}\left(\frac{1}{2}\right)^{n} ,\, \textrm{if}\,z \neq\frac{n}{2}.
\end{align*}

If  $n$ is even: $$\mathds{P}\left(n\tau = \frac{n}{2}\right)=\mathds{P}\left(n\tau_{\sigma_{1}}= \frac{n}{2},n\tau_{\sigma_{2}}= \frac{n}{2}\right)=\binom{n}{n/2}\left(\frac{1}{2}\right)^{n}$$
So we can conclude that   
	$$\mathds{P}(n\tau = z)= \left\{
	       \begin{array}{ll}
		 	\binom{n}{n/2}\left(\frac{1}{2}\right)^{n} & \mathrm{if}\, z=\frac{n}{2}\, \mathrm{and\ } n \mathrm{\ } \mathrm{is\ } \mathrm{even\ } \\ \\
		  2\binom{n}{z}\left(\frac{1}{2}\right)^{n}     & \mathrm{otherwise} \\
		    \end{array}
	     \right.$$

Then, the distribution function of $n\tau$ is:\\
\begin{align*}
F_{n\tau}(z)= & \mathds{P}(n\tau\leq z)=\mathds{P}(\min\left\{n\tau_{\sigma_{1}},n\tau_{\sigma_{1}}\right\}\leq z)\\
=& 1-\mathds{P}(\min\left\{n\tau_{\sigma_{1}},n\tau_{\sigma_{1}}\right\}\geq z+1)\\
=& 1-\mathds{P}(n\tau_{\sigma_{1}}\geq z+1,n\tau_{\sigma_{2}}\geq z+1)\\
=& \mathds{P}(n\tau \leq z)= \left\{
	       \begin{array}{ll}
		 \sum \limits^{z}_{i=0} 2\binom{n}{i} \left(\frac{1}{2}\right)^{n}= 2B_{n,1/2}(z)	& \mathrm{if\ } z+1\leq \frac{n}{2}\\
		  1      & \mathrm{if} z>\frac{n}{2} \\
		    \end{array}
	     \right.
\end{align*}

\paragraph{Proof of Proposition 4}

\begin{itemize}
\item Based on the table of joint probability, if $n$ is even\\ $\mathds{E}(n\tau)=\sum \limits^{n/2-1}_{k=0}2 \left(\frac{1}{2}\right)^{n}k\binom{n}{k}+\frac{n}{2} \left(\frac{1}{2}\right)^{n}\binom{n}{n/2}$.\\ The first term is:

$2 \left(\frac{1}{2}\right)^{n}\sum \limits^{n/2-1}_{k=1}k\binom{n}{k}=\left(\frac{1}{2}\right)^{n-1}\sum \limits^{n/2-1}_{k=1}kn\frac{\binom{n-1}{k-1}}{k}=\left(\frac{1}{2}\right)^{n-1}n\sum \limits^{n/2-1}_{k=1}\binom{n-1}{k-1}$

$=\left(\frac{1}{2}\right)^{n-1}n\sum \limits^{n/2-2}_{k=0}\binom{n-1}{k}=\left(\frac{1}{2}\right)^{n-1}n\left(\frac{2^{n-1}}{2}-\binom{n-1}{n/2-1}\right)$, 

as  $\sum \limits^{n-1}_{k=0}\binom{n-1}{k}=\sum \limits^{n/2-1}_{k=0}\binom{n-1}{k}+\sum \limits^{n-1}_{k=n/2}\binom{n-1}{k}=2^{n-1}$ 

and $\sum \limits^{n/2-1}_{k=0}\binom{n-1}{k}=\sum \limits^{n-1}_{k=n/2}\binom{n-1}{k}=\frac{2^{n-1}}{2}$. \\
We have now,\\
$\mathds{E}(n\tau)=\left(\frac{1}{2}\right)^{n-1}n \left(\frac{2^{n-1}}{2}-\binom{n-1}{n/2-1}\right)+\frac{n}{2} \left(\frac{1}{2}\right)^{n}\binom{n}{n/2}=$

$\frac{n}{2}- \left(\frac{1}{2}\right)^{n-1}n\binom{n-1}{n/2-1}+\frac{n}{2}\left(\frac{1}{2}\right)^{n}\binom{n}{n/2}=$

$\frac{n}{2}-\left(\frac{1}{2}\right)^{n-1}\binom{n}{n/2}\frac{n}{2}+\left(\frac{1}{2}\right)^{n}\binom{n}{n/2}\frac{n}{2}=\frac{n}{2}-\left(\frac{1}{2}\right)^{n}\binom{n}{n/2}\frac{n}{2}.$

\item Similarly if $n$ is odd:

$\mathds{E}(n\tau)=\sum \limits^{\frac{n-1}{2}}_{k=0}2\left(\frac{1}{2}\right)^{n}k\binom{n}{k}=\left(\frac{1}{2}\right)^{n-1}\sum \limits^{\frac{n-1}{2}}_{k=0}k\binom{n}{k}$

$=\left(\frac{1}{2}\right)^{n-1}\sum \limits^{\frac{n-1}{2}}_{k=0}n\binom{n-1}{k-1}=\left(\frac{1}{2}\right)^{n-1}n\sum \limits^{\frac{n-1}{2}-1}_{k=0}\binom{n-1}{k}$

$=\left(\frac{1}{2}\right)^{n-1}n\left[2^{n-2}-\frac{1}{2}\binom{n-1}{\frac{n-1}{2}}\right],$ as 

$\sum \limits^{n-1}_{k=0}\binom{n-1}{k}=\sum \limits^{\frac{n-1}{2}-1}_{k=0}\binom{n-1}{k}+\binom{n-1}{\frac{n-1}{2}}+\sum \limits^{n-1}_{k=\frac{n-1}{2}+1}\binom{n-1}{k}=$

$2\sum \limits^{\frac{n-1}{2}-1}_{k=0}\binom{n-1}{k}+\binom{n-1}{\frac{n-1}{2}}=2^{n-1}$ Then: $\sum \limits^{\frac{n-1}{2}-1}_{k=0}\binom{n-1}{k}=2^{n-2}-\frac{1}{2}\binom{n-1}{\frac{n-1}{2}}$.
\end{itemize}

Hence, 
	 \[\Rightarrow  
	\mathds{E}(n\tau)= \left\{
	       \begin{array}{ll}
		 \frac{n}{2}-(\frac{1}{2})^{n}\binom{n}{n/2}\frac{n}{2}	& \mathrm{if\ } n \,\, \mathrm{is}\,\,\mathrm{even\ }   \\
		  \frac{n}{2}-(\frac{1}{2})^{n}\binom{n-1}{\frac{n-1}{2}}n & \mathrm{if\ } n \,\, \mathrm{is }\,\, \mathrm{odd\ }  \\
		    \end{array}
	     \right.\]

\section*{Acknowledgments}
We thank the \emph{LIA-IFUM}, the \emph{ANII} -Uruguay and Campus France for their financial support.

\bibliographystyle{model2-names}
\bibliography{bibliografia}

\end{document}